\documentclass[]{spie}  

\usepackage{graphicx}
\usepackage[colorlinks=true, allcolors=blue]{hyperref}
\usepackage{aasmacros}

\title{The Lowell Observatory Solar Telescope: A fiber feed into the EXtreme PREcision Spectrometer}

\author[a]{Joe Llama}
\author[b]{Lily L. Zhao}
\author[c]{John M. Brewer}

\author[d]{Andrew Szymkowiak}
\author[d]{Debra A. Fischer}
\author[a]{Michael Collins}
\author[a]{Jake Tiegs}
\author[a]{Frank Cornelius}
\affil[a]{Lowell Observatory, 1400 W. Mars Hill Rd. Flagstaff, AZ. 86001. USA}
\affil[b]{Center for Computational Astrophysics, Flatiron Institute, Simons Foundation, 162 Fifth Avenue, New York, NY 10010, USA}
\affil[c]{Department of Physics and Astronomy, San Francisco State University, 1600 Holloway Ave, San Francisco, CA 94132, USA}
\affil[d]{Department of Astronomy, Yale University, Kline Tower, New Haven, CT 06511, USA}

\authorinfo{Further author information: (Send correspondence to J. Llama)\\J. Llama: E-mail: joe.llama@lowell.edu}

\begin{document} 
\maketitle

\begin{abstract}
The signal induced by a temperate, terrestrial planet orbiting a Sun-like star is an order of magnitude smaller than the host stars' intrinsic variability. Understanding stellar activity is, therefore, a fundamental obstacle in confirming the smallest exoplanets. We present the Lowell Observatory Solar Telescope (LOST), a solar feed for the EXtreme PREcision Spectrometer (EXPRES) at the 4.3-m Lowell Discovery Telescope (LDT). EXPRES is one of the newest high-resolution spectrographs that accurately measure extreme radial velocity. With LOST/EXPRES, we observe disk-integrated sunlight autonomously throughout the day. In clear conditions, we achieve a $R\sim137,500$ optical spectrum of the Sun with a signal-to-noise of 500 in $\sim150$s. Data is reduced using the standard EXPRES pipeline with minimal modification to ensure the data are comparable to the observations of other stars with the LDT. During the first three years of operation, we find a daily RMS of 71 cm/s. Additionally, having two EPRV spectrometers located in Arizona gives us an unprecedented opportunity to benchmark the performance of these planet-finders. We find a RMS of just 55 cm/s when comparing data taken simultaneously with EXPRES and NEID. 


\end{abstract}


\section{Introduction}\label{sec:intro}  
Since the detection of 51 Pegasi b in 1995, astronomers have confirmed over 5,000 exoplanets. Despite the incredible diversity and variety of newly discovered planets, the detection of a temperate Earth-mass exoplanet orbiting a star similar to the Sun remains, as yet, elusive ~\citenum{fischer2016}. One of the most promising methods to detect such a planet is radial velocity (RV).  When a planet orbits its host star, the gravitational pull of the planet induces periodic red- and blue-shifts in the host star's spectrum. The RV induced by a sun-like star due to a planet like Earth is just 9 cm/s. There are now several Extreme Precision Radial Velocity (EPRV) spectrographs that are routinely detecting planets with sub-m/s amplitudes \citenum{pepe2000,cosentino2012,jurgenson2016,schwab2016,seifahrt2018,pepe2021,crass2021}. 

Despite these advances in instrumentation, changes on the stellar surface result in time-varying spectral variations that can be misinterpreted as a change in the center of mass of the host star. These variations can be several ms$^{-1}$ in amplitude, an order of magnitude larger than an earth-mass exoplanet orbiting in the habitable zone of a sun-like star. This further complicates the challenges faced by EPRV planet detection whereby several different phenomena occurring at timescales of minutes to years compound to obscure the signal from an orbiting planet. On the shortest timescale, acoustic oscillations (such as p-modes) occur on a time scale of a few minutes ~\citenum{bouchy2005,kjeldsen2005,arentoft2008,dumusque2011a,chaplin2019}. Convective granulation ~\citenum{dravins1982,kjeldsen1995,lindegreen2003,dumusque2011a,meunier2015,cegla2018,lanza2019} and super-granulation occur on a few hours-days time scales ~\citenum{rincon2018,meunier2019}. On the time scale of the stellar rotation period (days-months), surface features such as starspots, faculae, and plage can rotate in and out of view changing the stellar rotational velocity profile and suppress convective blueshift ~\citenum{saar1997,hatzes2002,saar2003,desort2007,huelamo2008,dumusque2011b,jeffers2013,roettenbacher2022,haywood2016}. Finally, on the timescale of years-to-decades are stellar magnetic cycles. These increases and decreases in the levels of stellar activity are particularly problematic in the search for an actual Earth-analog exoplanet since they occur on the same time scale as the orbital period of the planet \citenum{meunier2010,egeland2017,luhn2022}

Much of our current understanding of RV variability induced by stellar activity comes from studies of our own Sun, such as systematic RV campaigns of the integrated whole-disk image of the Sun in order to mimic observations of a distant, point-like star with the HARPS spectrograph \citenum{haywood2014}. These early campaigns observed sunlight reflected off the bright asteroid 4/Vesta. Using disk-resolved images from the Helioseismic Magnetic Imager (HMI) onboard NASA's Solar Dynamics Observatory (SDO) ~\citenum{haywood2014} identified the individual surface features responsible for the solar RV variations, directly establishing that suppression of convective blueshift is the dominant process at play (confirming precursor work of ~\citenum{meunier2010}). These original studies highlighted the necessity of understanding stellar activity when searching for Earth-mass exoplanets and revealed the unique opportunity that the Sun provides.

Since the initial attempts to study the sun-as-a-star through reflected light with planet-finding spectrographs, many instruments have adopted a dedicated solar feed. These small telescopes observe the Sun like a star by integrating the entire solar disk, mimicking the point-source observations of other stars. These telescopes are relatively inexpensive to commission since they only require a small aperture to reach a high signal-to-noise ratio. Additionally, they can be used continuously throughout the day when there is zero demand for spectrometer usage for other science programs. The first such dedicated solar feed was installed on HARPS-N ~\citenum{phillips2016} and has led to a flurry of solar feeds being commissioned on EPRV spectrographs \citenum{lin2022,claudi2018,leite2022,rubenzahl2023}.

Here we present the commissioning and first four years of data from the Lowell Observatory Solar Telescope (LOST), the solar feed to the EXtreme PREcision Spectrometer (EXPRES \citenum{jurgenson2016}). In Section \ref{sec:instrument}, we present the optical and initial telescope designs. We also detail an upgrade performed to the telescope approximately two years into operation. Section \ref{sec:pipeline} discusses the modifications to the existing data reduction pipeline necessary for the solar telescope. In this Section, we also discuss the steps taken to assess the quality of each observation since the telescope is fully automated. We also discuss the offset at solar noon in the data that we attribute to the telescope performing a meridian flip that prompted the upgrade to the initial telescope design. Section \ref{sec:data} presents the first four years of data recorded with EXPRES/LOST as the Sun moves from cycle-minimum toward cycle-maximum. Finally, in Section \ref{sec:discussion}, we compare data obtained with LOST/EXPRES and the NEID solar feed (\citenum{lin2022}) taken simultaneously .

\section{LOST: The Lowell Observatory Solar Telescope}\label{sec:instrument}
\begin{figure}[t!]
    \centering
    \includegraphics[width=0.45\textwidth]{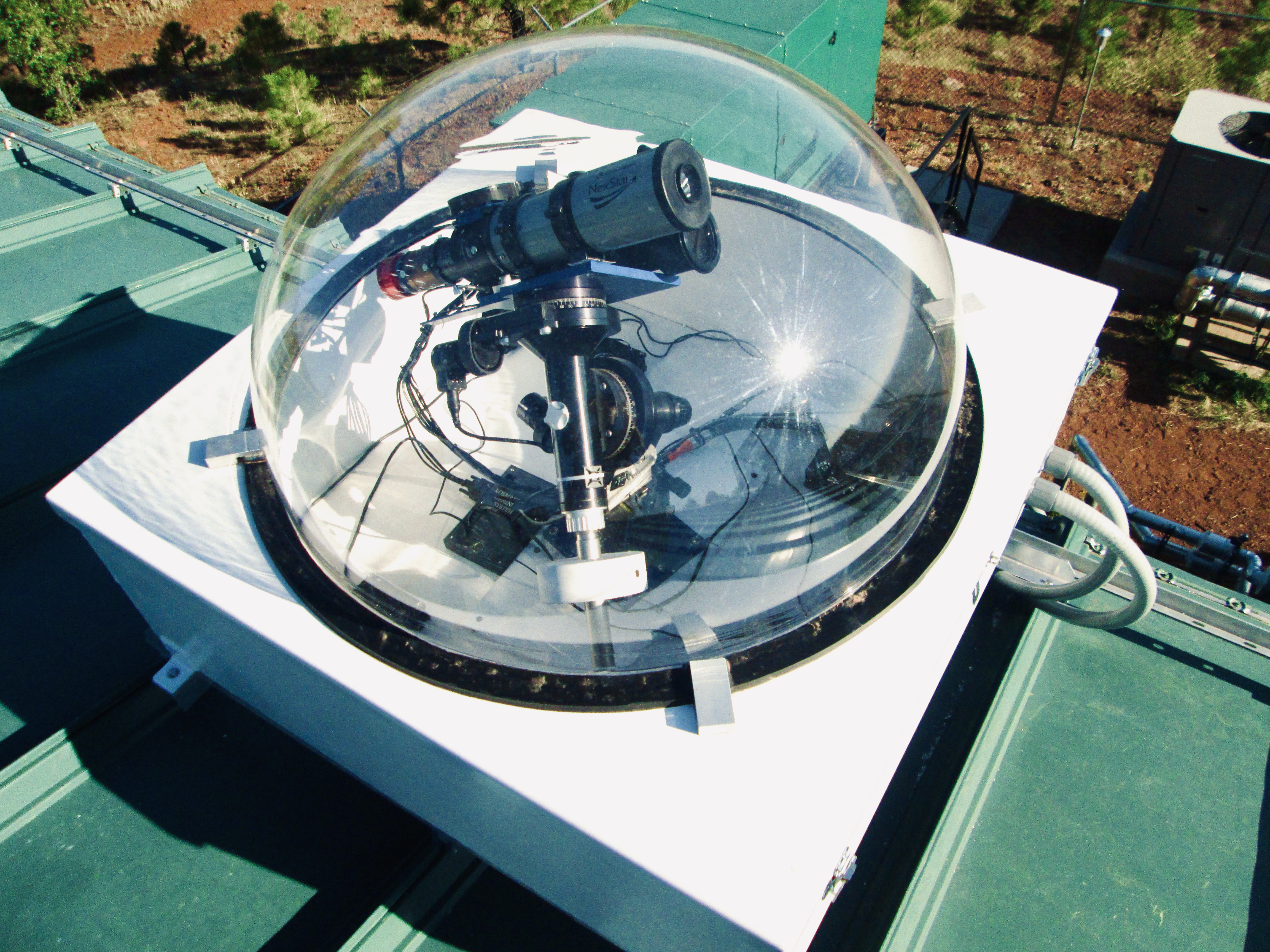}
    \includegraphics[width=0.45\textwidth]{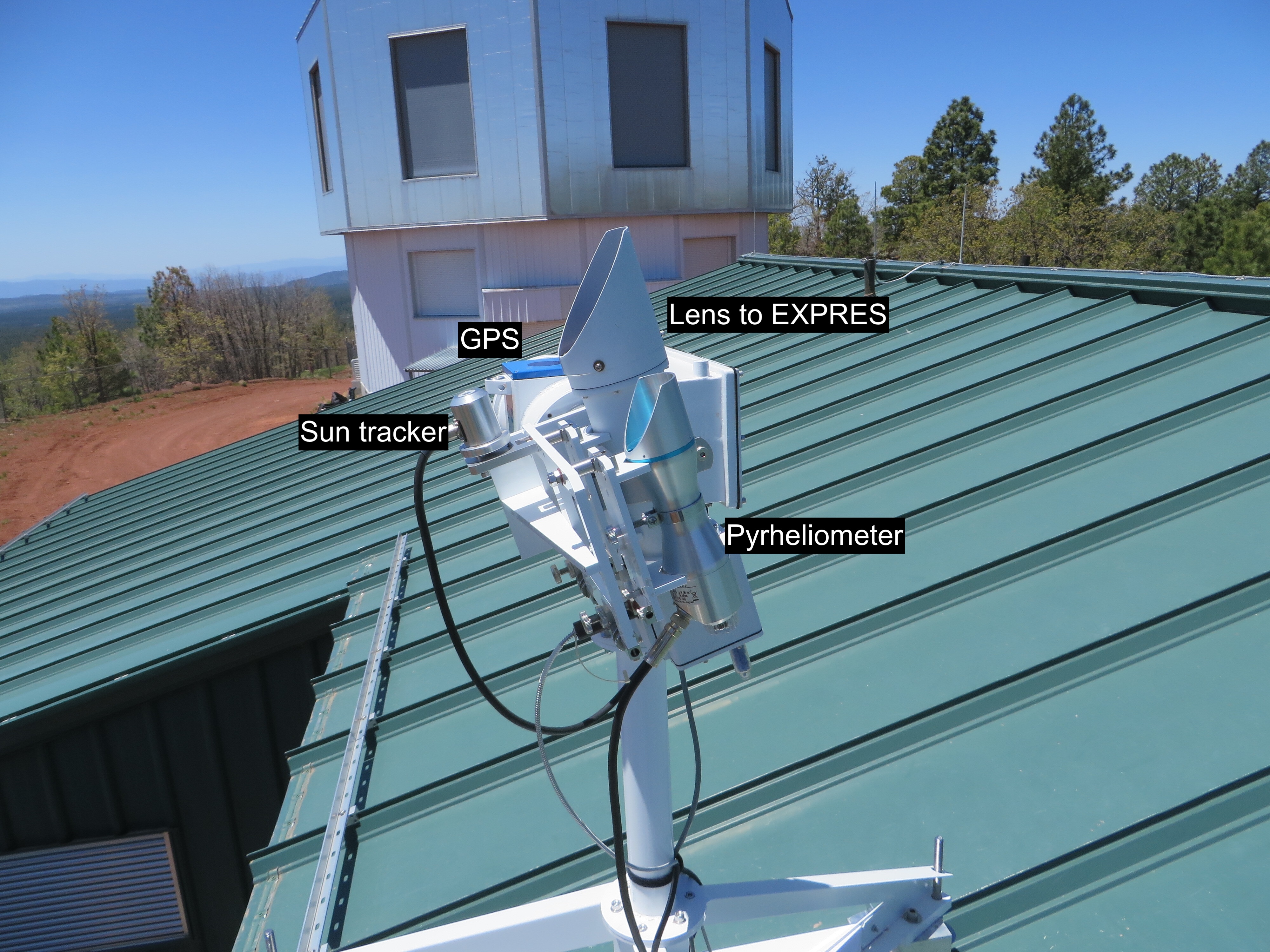}    
    \caption{{\bf Left:} The original design of the Lowell Observatory Solar Telescope (LOST). This telescope is a traditional equatorial mount that is housed in a weather-sealed enclosure. {\bf Right:} The upgraded design of LOST features a dedicated solar tracker from EKO systems and a custom-designed weather-sealed lens assembly. Both telescopes were mounted 
    on the roof of the auxiliary building at the Lowell Discovery Telescope site in Happy Jack, AZ, and fiber-fed into the EXtreme PREcision Spectrograph (EXPRES).  \label{fig:solar_telescope}}
\end{figure}
    The EXtreme PREcision Spectrometer (EXPRES; ~\citenum{jurgenson2016}) is a fiber-fed, stabilized spectrograph designed for EPRV detection of small exoplanets. It was commissioned at the 4.3-m Lowell Discovery Telescope (LDT; ~\citenum{levine2012}) in 2019 and is capable of long-term instrumental precision of $\sim30$ cm/s \citenum{brewer2020}. The Lowell Observatory Solar Telescope (LOST) began development in late 2019, approximately one year after the final commissioning of EXPRES. EXPRES was designed with a port for a solar feed from the beginning ~\citenum{jurgenson2016}; however, the solar feed itself was not part of the original funding for EXPRES. 

    At the highest level, the solar telescope consists of a small lens that feeds sunlight into an integrating sphere that spatially scrambles the solar image. This essentially turns the resolved disk of the Sun into a disk-integrated source like the stars observed at night. The integrating sphere is attached to a fiber that sends sunlight into EXPRES.

\subsection{Optics}
    The primary optic is a 75mm diameter achromatic doublet with a 200mm focal length from Edmund Optics (part no. 88-596). The optic was chosen as it had the best overlap with the EXPRES bandpass ($380-750$ nm). 
    The lens is housed in a $\sim200$ mm tube, at the focus of which is the entrance to a 2-in integrating sphere (ThorLabs part no. IS200/2P3). This integrating sphere has a heritage in other solar feeds and was chosen because of its high heat tolerance (for obvious reasons), high throughput, and relatively uniform response across the EXPRES wavelength range. The focus of the incoming beam is placed at the entry of the integrating sphere to ensure that all wavelengths of sunlight are captured in the integrating sphere. The opening into the integrating sphere is 11.5mm, and the solar image at the focal distance of the achromatic lens is 1.75mm. Having the solar image be much smaller than the input diameter minimizes the risk that not all the light from the solar image enters the integrating sphere. 

    The Thorlabs integrating sphere accepts standard fiber attachments through FC/PC connectors. The fiber to feed the sunlight from LOST into EXPRES was ordered from FiberTechOptica. We chose a broadband solarization-resistant optical fiber with a response between 200-2100 nm (part no. SBB200/220PI). The fiber is jacketed in a weather-resistant coating. The total length of the fiber is 80m, which is the approximate distance between the solar telescope and EXPRES. The cable was run through an underground conduit between the auxiliary building and the 4.3m Lowell Discovery Telescope, where EXPRES is housed. The fiber enters the spectrograph through a shutter designed explicitly for the solar feed. From there, the light travels up the calibration fiber to the EXPRES front-end module on the LDT and back to the spectrograph along the science fiber. This ensures sunlight travels along the same light path as the nightly science data.  

\subsection{LOST-1: November 2020 - September 2023}
\begin{figure}
\centering
    \includegraphics[width=\textwidth]{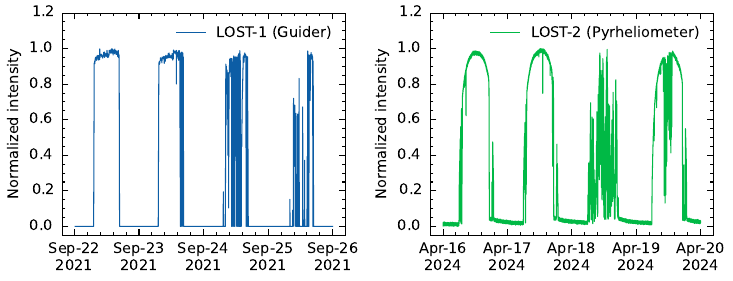}
    \caption{We record the intensity of the Sun using independent devices for both LOST-1 and LOST-2. For LOST-1 (left), we use data from the Hinode solar tracker. For LOST-2 (right), the pyrheliometer accurately measures the solar intensity every 1-s. In the LOST-1 data, on clear days, the solar intensity appears to ramp up throughout the day (rather than rise and then fall before/after solar noon), which we attribute to the poor quality of the acrylic dome of the enclosure (see Section \ref{sec:jump} for more information).  \label{fig:guider_intensity}}
\end{figure}
    The original design of the solar telescope was based on the HARPS-N solar feed. The lens assembly was mounted onto a Losmandy G11 equatorial mount. This mount was chosen due to its heritage within the amateur astronomy community and its ability to be remotely controlled using serial commands. The payload is far below the maximum weight allowed by the mount, and we custom-built a counterweight to ensure maximum stability of the telescope.   

    Since the telescope mount is not weather resistant, we followed the design of the HARPS-N solar feed and housed the telescope in a custom-built enclosure (Figure \ref{fig:solar_telescope}; left panel). The size of the Losmandy G11 required us to use a larger enclosure than the solar tracker used for HARPS-N. The enclosure was fabricated out of steel and included a 40-inch acrylic dome. The enclosure was mounted to the roof of the LDT auxiliary building. We ran a conduit to route electronics and fiber into the auxiliary building. The dome is fully weather-sealed using rubber gaskets and can be levered open to allow access for maintenance. 

    Since the telescope is an amateur-style mount, it must be commanded to point to the Sun and begin tracking. The Losmandy G11 has a well-documented list of serial commands that can be sent to the mount to command it and receive status information. We used \texttt{Python} and {\tt pyserial} to interact with the mount fully autonomously. The software controls the telescope using {\tt serial} commands and communicates the system's status using {\tt websockets}. At the beginning of each day, we use the \texttt{astropy/get\_sun} routine to determine the coordinates of the Sun, convert those into the \texttt{AltAz} frame using \texttt{Astropy/EarthLocation} to obtain the RA/DEC, and airmass of the Sun. We compute the time at which the Sun will rise above $15^\circ$ and then use \texttt{pyserial} to send a command to the mount to slew the telescope to the RA/DEC of the Sun. Through testing, we found the mount to be accurate to within a few degrees; however, the addition of the solar guider ensures stable tracking to a few arc minutes. Once the telescope has reached the destination coordinates, we begin tracking and activate the guider again using \texttt{pyserial}.

    Precise guiding is fundamental to ensure that the entire disk of the Sun is projected into the integrating sphere; any loss of the solar limb would result in a large RV offset in the resultant data. We minimize this risk by ensuring that the field-of-view of LOST is much larger than the solar disk. We purchased a guider specifically designed to track the Sun from Hinode Systems.  The guider monitors the solar intensity and sends commands to the mount to ensure the telescope is aligned and the solar image remains centered in the field of view. The guider consists of a wide-field camera that measures the intensity of the Sun and sends commands directly to the telescope mount to align the Sun and ensure the solar disk remains centered. When the solar intensity goes below a certain threshold (e.g., due to clouds), guiding stops, and the telescope continues to track until the Sun reappears. The guider can also be queried using serial commands. Our control software can enable guiding during the day and disable tracking at night. While tracking, the software queries the guider and saves the solar intensity every 10s. This measurement provides an independent estimate of the solar intensity that we can compare with the EXPRES exposure meter. The left panel of Figure \ref{fig:guider_intensity} shows four days of normalized intensity of the Sun as recorded by the Hinode solar tracker. The first two days were clear with minimal cloud coverage, while the last two days had heavy cloud obscuration.  

  \subsection{LOST-2: September 2023-Present}
    
    In September 2023, we upgraded the original telescope mount to a dedicated solar tracker. This upgrade was prompted by the discovery of a jump in the radial velocity data with LOST-1/EXPRES after the telescope performed its meridian flip. Section \ref{sec:jump} will discuss this discontinuity in more detail.  While we can correct for the discontinuity using regression techniques, a more satisfactory solution was to replace the mount with one that does not require a weather-sealed dome. 
    
    The updated design is based on the NEID solar feed, which, rather than being based on a traditional telescope mount, uses a dedicated solar tracker from EKO systems that is entirely weather-sealed \citenum{lin2022}. We purchased the STR-32G Sun tracker, which is EKO's high torque and load option that provides the most precision in extreme conditions (such as those experienced in Northern Arizona during the winter), and the MS-57 Pyrheliometer to provide accurate measurements of the solar intensity. 

    The right image of Figure \ref{fig:solar_telescope} shows the payload for the upgraded solar telescope that uses an EKO solar tracker. Controlling the EKO solar tracker in LOST-2 is much more automated than the Losmandy mount used for LOST-1. Since the tracker is dedicated to solar observation, it is automatically programmed out-of-the-box to slew to the Sun once it is above the horizon and begin tracking. The mount uses a GPS receiver to determine its location and date/time. The mount has a sun tracker (similar to the Hinode tracker) that sends offsets to the mount to ensure the Sun remains centered.
    
    Similarly, when a cloud occults the Sun, the mount tracks without guiding and resumes offset correction when the Sun reappears.  Additionally, the mount parks itself in the evening once the Sun goes below the horizon. The lens to EXPRES is the same optics used in LOST-1, but it is now housed in a weather-proof enclosure (identical to the NEID lens assembly) since the acrylic dome no longer protects it. 

    In addition to the sun tracker, we also installed the EKO MS-57 pyrheliometer. The pyrheliometer accurately measures the solar intensity at a 1 s cadence throughout the day. The right panel of Figure \ref{fig:guider_intensity} shows four days of normalized pyrheliometer data. The first two days (Apr 16-17) were largely cloud-free; however, Apr 18 and 19 suffered considerable loss in intensity due to cloud coverage. Unlike LOST-1, where the intensity of the Sun uniformly increases during the day (due to the acrylic dome), the recorded intensity with LOST-2 is symmetric around solar noon, which is expected as the intensity of the Sun should correlate with altitude.

\subsection{Observing strategy}
    EXPRES solar observing begins at 07:30 with a series of ``Beginning of Day'' calibrations. These calibrations are identical to our ``Beginning of Night''  calibrations and consist of thirty dark exposures (0s) and thirty flats (20s). Combined with the $\sim$40s read-out time, the calibration sequence takes approximately 1-hour to complete. The solar telescope control system triggers observations with EXPRES using {\tt websockets}, and the system can listen to the instrument's current status and provide real-time information through a web interface designed for monitoring the status of the solar telescope. After each exposure is complete, the system triggers the subsequent exposure. As with our nightly EPRV observations, we perform wavelength calibration every 30 minutes. Our wavelength calibration consists of a 30s Thorium Argon (ThAr) followed by a 15s Laser Frequency Comb (LFC) observation, which, when combined with readout, lasts approximately 2 minutes. 

    The current generation EPRV instruments, including EXPRES, use a dynamic exposure meter. The dynamic exposure meter in EXPRES is a $R\sim100$ spectrograph that records the incident stellar flux over a similar bandpass to EXPRES at 1s cadence ~\citenum{petersburg2020}. The flux recorded by the exposure meter primarily serves two purposes: First, the time series allows us to flux-weight the barycentric correction in the reduction step. Second,    
    it allows for real-time monitoring of the signal-to-noise (SNR) during an exposure, which allows us to terminate an exposure when the desired SNR has been reached. This not only allows for uniformity in the RV time series we obtain, it also maximizes our efficiency on-sky.  The solar telescope control triggers an exposure to end once a SNR of 500 has been reached, or 600s, whichever occurs first. 600s was chosen because it is approximately $3\times$ longer than the typical exposure time required in clear conditions. 
    
    Since LOST-1 was an equatorial mount, once the Sun transits the meridian, it must perform a meridian flip to continue tracking. Any observation taken while the telescope is slewing is flagged and automatically rejected. Once the Sun reaches an altitude of $15^\circ$, the solar exposures are terminated, and the telescope is slewed to the park position for the night. As a quality-of-life implementation, the solar telescope control software checks if EXPRES is scheduled for evening observations, and if it is, automatically obtains ``Beginning of night'' calibrations (30 dark, and 30 flat frames). 

    As with the nightly EXPRES data, all the files are stored on a Synology NAS housed at the telescope that was initially purchased to handle the large amount of data the solar telescope would produce. The NAS has 12x16 TB Seagate EXOS HDDs in a RAID configuration.  This data is then copied periodically throughout the day to an identical system at Lowell Observatory's main campus in Flagstaff.

\section{Data reduction}\label{sec:pipeline}
\subsection{Modifications to the reduction pipeline}
    At the end of each day, the data are reduced using a modified version of the EXPRES nightly pipeline ~\citenum{petersburg2020}. The goal of the solar telescope is to obtain solar radial velocities in a fashion that is as close to those we obtain at night. As a result, very minimal modifications had to be made to the pipeline. Here, we focus our discussion on the changes to the pipeline that relate to the solar data and refer the reader to ~\citenum{petersburg2020} for a complete description of the EXPRES pipeline. 
    
    The majority of the modifications for the pipeline were made to the wavelength determination. In particular, the treatment of solar observations requires a different barycentric correction to other stars (which are outside the solar system). The EXPRES pipeline uses \texttt{barycorrpy} to apply this correction ~\citenum{kanodia2018}. Given the recent focus on stellar activity studies with EPRV spectrographs, the latest version of \texttt{barycorrpy} introduces a standardized approach for applying an analogous barycentric correction to solar observations obtained with solar telescopes attached to spectrographs such as EXPRES, HARPS, and NEID ~\citenum{kanodia2018}. As with the nightly EPRV observations, the barycentric correction is flux weighted by the counts-per-second recorded using the exposure meter.

\begin{figure*}[!t]
    \centering
    \includegraphics[width=\textwidth]{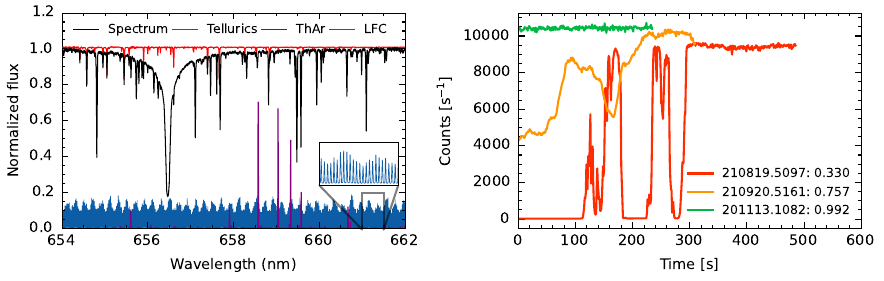}
    \caption{{\bf Left:} A typical EXPRES solar observation zoomed-in to the H$-\alpha$ line at 656-nm. The continuum-normalized spectrum is shown in black, and the telluric model is shown in red. Below the spectrum are the two wavelength calibration sources: Thorium Argon (purple) is used for the initial wavelength, and then the densely sampled Laser Frequency Comb (blue) is used to refine the wavelength solution and provide sub-m/s precision. {\bf Right:} Since the solar data is obtained autonomously, having a stringent and robust metric for assessing the data quality is fundamental. For LOST/EXPRES, we use the EXPRES exposure meter data to produce a ``light curve'' of the observation and measure its duration and variability. Here are three exposures with varying quality factors. We find a threshold of $>0.95$ to be a reasonable cut to reject solar RVs. }
    \label{fig:spectrum}
\end{figure*}    
    Refinement of the wavelength solution is done using {\tt Excalibur}. {\tt Excalibur ~\citenum{zhao2021}} is a hierarchical, non-parametric method that returns five times more precise wavelength solutions, i.e., the mapping of the wavelength of light as a function of detector position. {\tt Excalibur} utilizes all calibration images (ThAr and LFC) to construct a model of the accessible calibration space of an instrument, which is used to de-noise line positions and pinpoint the calibration state of the instrument at any time ~\citenum{zhao2021}. {\tt Excalibur} is run independently on the solar and nightly data as the respective data products are reduced and analyzed separately

    Superimposed on every observation taken with EXPRES is the signature of the Earth's atmosphere that must be removed from the spectrum before a radial velocity can be computed. Even in the optical, microtellurics can induce RV variability of 50 cms$^{-1}$. To remove tellurics, the EXPRES pipeline utilizes a self-calibrated empirical linear regression model to accurately model telluric lines in high-resolution spectra known as {\tt SELENITE} ~\citenum{leet2019}. 

    The left panel of Figure \ref{fig:spectrum} shows the resultant spectrum of a solar observation taken with EXPRES zoomed in on the H$\alpha$ line at 656.6 nm. The black line is the extracted spectrum of the star that has been continuum-normalized as part of the extraction. The wavelength calibration is shown in purple (ThAr) and blue (LFC). Figure \ref{fig:spectrum} also highlights the density of lines in the LFC as compared to the traditional ThAr calibration source. Finally, over-plotted in red is the telluric model for this exposure.

\subsection{Quality control}
Very quickly, in the routine operation of the solar feed, it became apparent we needed a metric to assess the quality of the observations, e.g., a robust method to evaluate whether observations were impacted by cloud or pointing issues. Various approaches to quality control of solar observations have been reported in the literature. The HARPS-N solar telescope uses a Gaussian Mixture Model to assess the quality of observations ~\citenum{colliercameron2019}. We analyze the dynamic exposure meter for LOST to monitor the solar intensity during the exposure. The exposure meter records the counter-per-second (cps) every second over the EXPRES bandpass with a resolution, $R\sim100$. This data is extracted and binned into eight wavelength channels as part of the pipeline extraction. 

As a first attempt at assigning a ``quality factor'' to the solar observations, we further bin the exposure meter data into a single channel and produce a ``light curve'' with a 1-s cadence of solar intensity for each exposure. We then compute the metric,
\begin{equation}
\rm{qf} = 1 - \frac{\rm{std}(\rm{cps})}{\rm{med}({\rm{cps}})},
\end{equation}
where std(cps) is the standard deviation of the counts-per-second in the ``light curve'' and med(cps) is the median. 
The right panel of Figure \ref{fig:spectrum} shows three examples of ``light curves'', i.e., time vs solar intensity derived from the EXPRES exposure meter, and the corresponding quality factor is shown in the legend. The first observation (210819.5097) is impacted by heavy temporal cloud occultation that causes the counts-per-second to fluctuate. The second observation (210920.5161) is likely partially obscured by a thin cloud moving across the solar disk, as the intensity does not fall to zero. Finally, the third observation (201113.1082) is taken under clear conditions. In all cases, the exposure is set to terminate when the SNR reaches 500 or after 600-s, whichever occurs first. Despite two of these three observations being affected by weather, all three terminate before the 600-s cut-off. This emphasizes the need for stringent quality control of solar observations.

\subsection{Offset due to meridian flip}\label{sec:jump}
\begin{figure}[h]
    \centering
    \includegraphics[width=\textwidth]{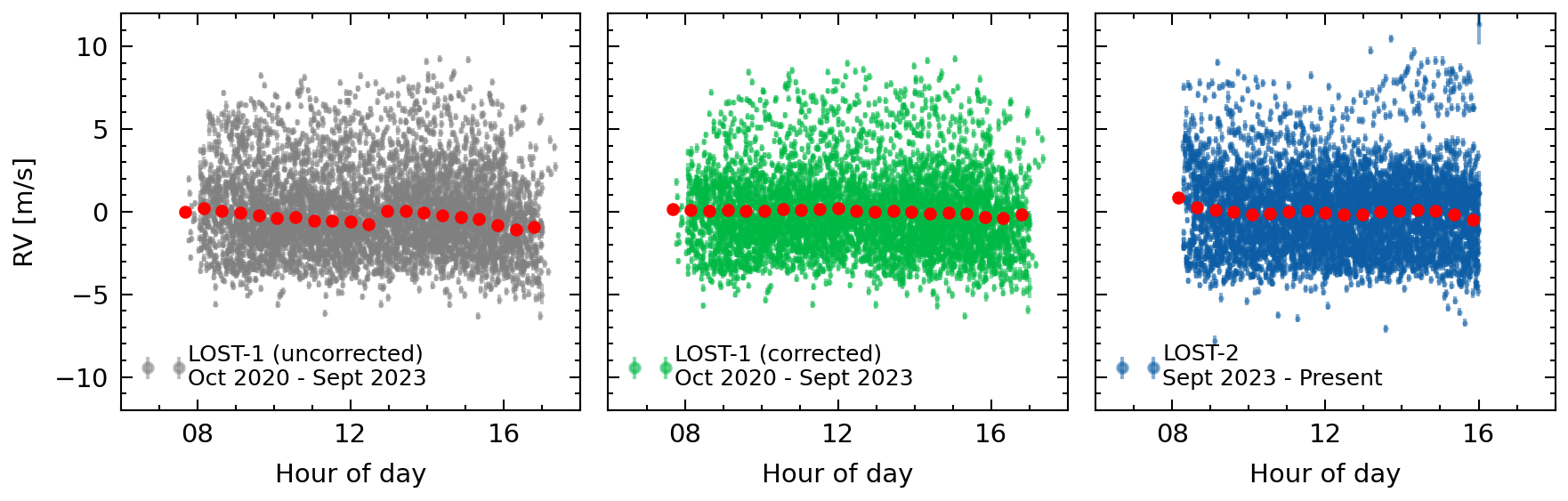}
    \caption{{\bf Left:} Radial velocities from LOST-1 phased to the same time of day demonstrate a jump near solar noon when the telescope performs a meridian flip, which we attribute to the poor optical quality of the acrylic dome. {\bf Center:} We perform a regression against time from the meridian flip to remove this trend from the LOST-1 data. {\bf Right:} LOST-2 removed the need for a dome. Additionally, the telescope does not require a meridian flip to track the Sun throughout the day. We do not see a jump in the data with the upgraded system. }
    \label{fig:jump_correction}
\end{figure}
    As discussed in Section \ref{sec:instrument}, LOST-1 used an equatorial mount housed in a weatherproof enclosure and observed the Sun through a large acrylic dome. As a result, the telescope had to perform a meridian flip at (or near) solar noon. Shown in the left panel of Figure \ref{fig:jump_correction} are all the RV's with a quality factor $>0.9$ phased to the same day (gray) with hourly binned shown in red taken with LOST-1. The RV time series has a discontinuity after the telescope performs the meridian flip. This jump is also seen in the guider data on clear days. Since the guider is external to EXPRES, we have attributed this flux increase to the acrylic dome's poor optical quality.
    
    The observations were linearly regressed against the time from when the telescope performed the meridian flip to correct for the discontinuity.  Given the sharp discontinuity during the flip, an offset for before and after the fit was also included in the model. The model, therefore, has three free parameters.  Regressing against time from flip also captures the effects of differential extinction.  The central panel of Figure \ref{fig:jump_correction} shows the results of applying the best-fit linear correction to the observations. The right panel of Figure \ref{fig:jump_correction} shows the RVs from LOST-2, which no longer performs a meridian flip.  To keep data consistent, LOST-2 data are still regressed against time of day to remove the effects of differential extinction.
\begin{figure}[t]
    \centering
    \includegraphics[width=\textwidth]{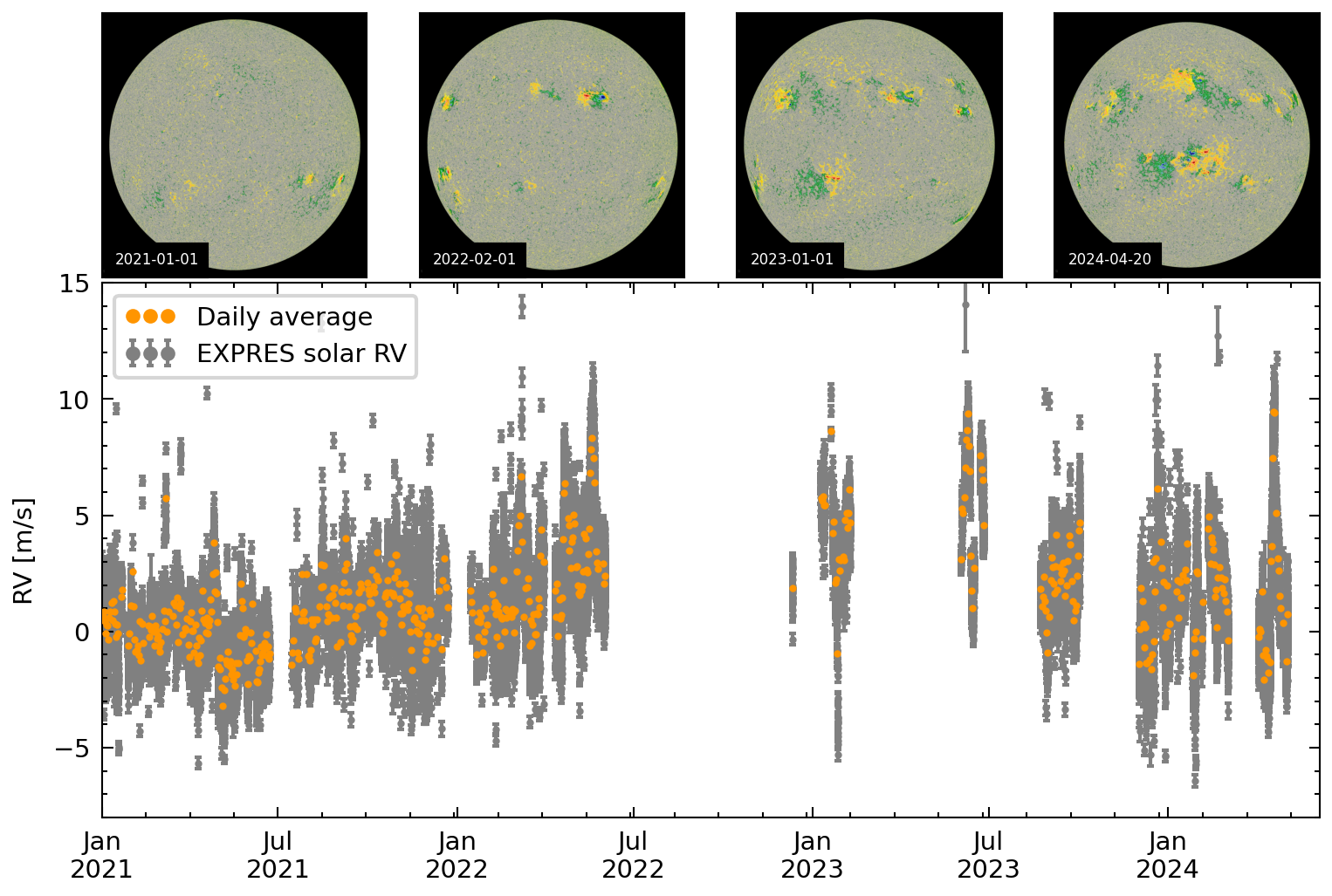}
    \caption{We have been observing the Sun with LOST/EXPRES since late 2020 and have collected over 40,000 good-quality RVs (gray points). Over plotted in orange is the daily average RV. We began observing the Sun during cycle minimum; since then, the level of magnetic activity on the Sun (top-row; SDO/HMI) has continued to increase as the solar cycle 25 progresses towards its maximum. }
    \label{fig:solar_rvs}
\end{figure}    
    The discovery of the offset prompted us to investigate various possibilities for an upgrade to the solar telescope. We considered upgrading the enclosure to a dome that could be opened during the day and when the weather permits, as well as a new type of mount that does not need to be protected from the elements. Ultimately, we chose the latter route given the success of the NEID solar feed, which uses a dedicated solar tracker from EKO systems. The new telescope was commissioned in September 2023, and the right panel of Figure \ref{fig:jump_correction} shows the RVs obtained since the upgrade was completed in September 2023. The jump at noon is no longer present because the EKO mount can continuously track the Sun throughout the day without performing a meridian flip. 
    

\section{Initial data results: Observing Solar Cycle 25 with LOST/EXPRES}\label{sec:data}
    We began operating the EXPRES solar telescope in late 2020 when the Sun was at the minimum of its 11-year cycle. Since then, we have collected  $\sim40,000$ radial velocities that meet our quality control threshold. During this time, the Sun has also increased in activity as Solar Cycle 25 moves towards its maximum (currently predicted to be in July 2025). The top row of Figure \ref{fig:solar_rvs} shows magnetograms of the Sun taken with the Helioseismic and Magnetic Imager (HMI; \citenum{scherrer2012}) onboard NASA's Solar Dynamics Observatory (SDO; \citenum{pesnell2012}). The images clearly show the increasing fraction of the solar surface covered in magnetically active regions as the solar cycle progresses. The lower panel of Figure \ref{fig:solar_rvs} shows the radial velocities that met our quality threshold $(>0.9)$ as gray points with a daily average over-plotted. The large gap in 2022 was due to an accidental break in our fiber that resulted in $\sim4$ months of downtime while we obtained a replacement. 

    From the 40,000 RVs that meet our quality control threshold, we find a median exposure time of $\sim180$-s for LOST-1 and 140-s for LOST-2. We attribute this increase in throughput to removing the acrylic dome in the upgrade from LOST-1 to LOST-2. We find that the daily RMS for days where we have more than ten high-quality observations is 71 cm/s.

\section{Conclusions and Prospects}\label{sec:discussion}
    We have commissioned the Lowell Observatory Solar Telescope (LOST), a solar feed into the EXtreme PREcision Spectrometer (EXPRES). LOST/EXPRES is one of the latest solar feeds into an EPRV spectrometer and joins several other instruments that observe our Sun during the day to help us mitigate the impact of stellar activity on exoplanet detection. 
    
    LOST/EXPRES is in Northern Arizona at the 4.3-m Lowell Discovery Telescope. In Southern Arizona, NASA and NSF's NEID spectrograph is installed on the 3.5-meter WIYN Telescope at Kitt Peak National Observatory, approximately 200 miles ($\sim320$-km) south of the LDT. Having two EPRV spectrographs so close to each other that they are both observing the Sun simultaneously provides a unique opportunity to benchmark the spectrographs' performance and check for any unseen issues in the data reduction pipelines. 
\begin{figure}[h]
    \centering
    \includegraphics[width=\textwidth]{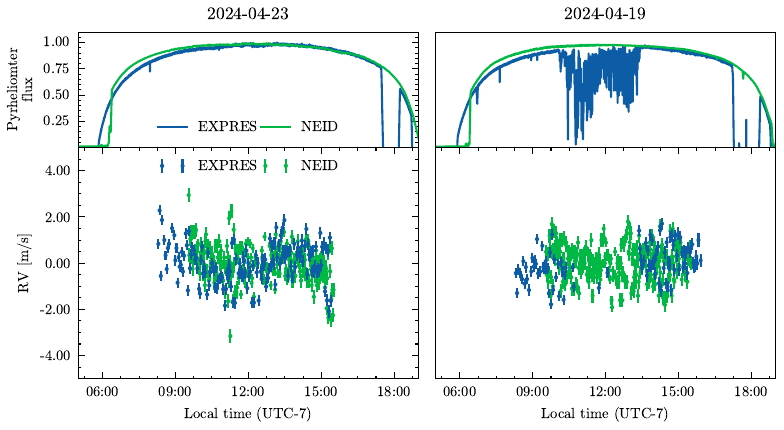}
    \caption{EXPRES and NEID are two of the newest EPRV spectrometers to have solar feeds measuring solar RVs. Having these two spectrometers co-located in Arizona offers an unprecedented opportunity to benchmark instrument performance and extend the coverage of solar RVs when one site is weathered out. On clear days (left), we find remarkable agreement between the instruments. Comparing data taken in the first 5-months of 2024, we find the RMS difference between EXPRES and NEID to be 55 cm/s. Additionally, multiple spectrographs allow us to maximize the coverage of the solar observations in the presence of bad weather at one site. The right panel shows the lack of RVs obtained with EXPRES when the pyrheliometer recorded a decrease in solar irradiance, yet the NEID spectrograph was able to continue obtaining RVs since the bad weather was localized to Northern Arizona. }
    \label{fig:expres_vs_neid}    
\end{figure}
    Figure \ref{fig:expres_vs_neid} shows two days of solar RVs measured by EXPRES and NEID. The top panel shows the data from the respective pyrheliometers, and the bottom panel shows the radial velocity. On clear days, We find remarkable agreement between the two spectrographs (e.g., left panel). The right panel highlights an advantage of having more than one solar feed; on days when the Sun is obscured at one site (e.g., due to bad weather) or the instrument isn't available for solar observing, the other instrument continues to collect data. Indeed, the Extreme Stellar-signals project (ESSP; \citenum{zhao2023}) was a coordinated effort to compare observations from four solar feeds: EXPRES, NEID, HARPS-N, and HARPS. While the time coverage between EXPRES and NEID is similar, the two instruments use a different cadence, and the observations are not synced at the start/end time. As a result, a direct comparison between the data requires binning onto a standard time frame. For the ESSP, the data were binned onto a standard time frame, and a comparison of 508 data points between EXPRES and NEID yielded an RMS scatter of 39-cm/s (Figure 7; \citenum{zhao2023}). At the time of the ESSP publication, we were still using LOST-1. After the upgrade to LOST-2, we repeated the comparison with approximately three months' worth of data. We compared RVs taken in the first 5 months of 2024 with NEID and LOST to asses the impact of upgrading the telescope mount. Overall, we see an RMS scatter of 55-cm/s, which is larger than the value reported in \citenum{zhao2023}. We attribute this increase in scatter to the more extended baseline ($\sim5$ months vs. 1 month in ESSP) for comparison. Additionally, the Sun is more active in the time frame analyzed here than the month of observations used for the ESSP analysis.

As new EPRV spectrographs are brought online, the addition of solar feeds offers the opportunity to study our own star with unprecedented cadence, coverage, and precision that isn't possible for other stars. These datasets allow us to compare pipelines and instrument performance and diagnose potential issues. Combining data from EPRV solar feeds at multiple longitudes (e.g., HARPS-N in Tenerife ~\citenum{phillips2016}, EXPRES/NEID in Arizona, and, most recently, SoCal on Mauna Kea ~\citenum{rubenzahl2023}) will further increase our temporal coverage of solar RVs and bring us closer and closer to determining the best methods for dissecting the signals of stellar activity from those of temperate, Earth-mass exoplanets.

\section*{Acknowledgements}
Lowell Observatory sits at the base of mountains sacred to tribes throughout the region. We honor their past, present, and future generations, who have lived here for millennia and will forever call this place home.

We acknowledge the generous support to purchase hardware for this work from the Mt. Cuba Astronomical Foundation, M. Collins, and Lowell Observatory. We are also grateful for telescope time support from Heising-Simons, which has enabled a vigorous research program at the LDT. 

J.L. Acknowledges support from NASA grant 80NSSC23K0040. 

\bibliography{solar} 

\begin{thebibliography}{10}

\bibitem{fischer2016}
{Fischer}, D.~A., {Anglada-Escude}, G., {Arriagada}, P., {Baluev}, R.~V., {Bean}, J.~L., {Bouchy}, F., {Buchhave}, L.~A., {Carroll}, T., {Chakraborty}, A., {Crepp}, J.~R., {Dawson}, R.~I., {Diddams}, S.~A., {Dumusque}, X., {Eastman}, J.~D., {Endl}, M., {Figueira}, P., {Ford}, E.~B., {Foreman-Mackey}, D., {Fournier}, P., {F{\H{u}}r{\'e}sz}, G., {Gaudi}, B.~S., {Gregory}, P.~C., {Grundahl}, F., {Hatzes}, A.~P., {H{\'e}brard}, G., {Herrero}, E., {Hogg}, D.~W., {Howard}, A.~W., {Johnson}, J.~A., {Jorden}, P., {Jurgenson}, C.~A., {Latham}, D.~W., {Laughlin}, G., {Loredo}, T.~J., {Lovis}, C., {Mahadevan}, S., {McCracken}, T.~M., {Pepe}, F., {Perez}, M., {Phillips}, D.~F., {Plavchan}, P.~P., {Prato}, L., {Quirrenbach}, A., {Reiners}, A., {Robertson}, P., {Santos}, N.~C., {Sawyer}, D., {Segransan}, D., {Sozzetti}, A., {Steinmetz}, T., {Szentgyorgyi}, A., {Udry}, S., {Valenti}, J.~A., {Wang}, S.~X., {Wittenmyer}, R.~A., and {Wright}, J.~T., ``{State of the Field: Extreme Precision Radial Velocities},'' {\em
  \pasp}~{\bf 128},  066001 (June 2016).

\bibitem{pepe2000}
{Pepe}, F., {Mayor}, M., {Delabre}, B., {Kohler}, D., {Lacroix}, D., {Queloz}, D., {Udry}, S., {Benz}, W., {Bertaux}, J.-L., and {Sivan}, J.-P., ``{HARPS: a new high-resolution spectrograph for the search of extrasolar planets},'' in [{\em Optical and IR Telescope Instrumentation and Detectors}{\nolinebreak\hspace{0.1em}]},  {Iye}, M. and {Moorwood}, A.~F., eds., {\em Society of Photo-Optical Instrumentation Engineers (SPIE) Conference Series} {\bf 4008},  582--592 (Aug. 2000).

\bibitem{cosentino2012}
{Cosentino}, R., {Lovis}, C., {Pepe}, F., {Collier Cameron}, A., {Latham}, D.~W., {Molinari}, E., {Udry}, S., {Bezawada}, N., {Black}, M., {Born}, A., {Buchschacher}, N., {Charbonneau}, D., {Figueira}, P., {Fleury}, M., {Galli}, A., {Gallie}, A., {Gao}, X., {Ghedina}, A., {Gonzalez}, C., {Gonzalez}, M., {Guerra}, J., {Henry}, D., {Horne}, K., {Hughes}, I., {Kelly}, D., {Lodi}, M., {Lunney}, D., {Maire}, C., {Mayor}, M., {Micela}, G., {Ordway}, M.~P., {Peacock}, J., {Phillips}, D., {Piotto}, G., {Pollacco}, D., {Queloz}, D., {Rice}, K., {Riverol}, C., {Riverol}, L., {San Juan}, J., {Sasselov}, D., {Segransan}, D., {Sozzetti}, A., {Sosnowska}, D., {Stobie}, B., {Szentgyorgyi}, A., {Vick}, A., and {Weber}, L., ``{Harps-N: the new planet hunter at TNG},'' in [{\em Ground-based and Airborne Instrumentation for Astronomy IV}{\nolinebreak\hspace{0.1em}]},  {McLean}, I.~S., {Ramsay}, S.~K., and {Takami}, H., eds., {\em Society of Photo-Optical Instrumentation Engineers (SPIE) Conference Series} {\bf 8446},  84461V
  (Sept. 2012).

\bibitem{jurgenson2016}
{Jurgenson}, C., {Fischer}, D., {McCracken}, T., {Sawyer}, D., {Szymkowiak}, A., {Davis}, A., {Muller}, G., and {Santoro}, F., ``{EXPRES: a next generation RV spectrograph in the search for earth-like worlds},'' in [{\em Ground-based and Airborne Instrumentation for Astronomy VI}{\nolinebreak\hspace{0.1em}]},  {Evans}, C.~J., {Simard}, L., and {Takami}, H., eds., {\em Society of Photo-Optical Instrumentation Engineers (SPIE) Conference Series} {\bf 9908},  99086T (Aug. 2016).

\bibitem{schwab2016}
{Schwab}, C., {Rakich}, A., {Gong}, Q., {Mahadevan}, S., {Halverson}, S.~P., {Roy}, A., {Terrien}, R.~C., {Robertson}, P.~M., {Hearty}, F.~R., {Levi}, E.~I., {Monson}, A.~J., {Wright}, J.~T., {McElwain}, M.~W., {Bender}, C.~F., {Blake}, C.~H., {St{\"u}rmer}, J., {Gurevich}, Y.~V., {Chakraborty}, A., and {Ramsey}, L.~W., ``{Design of NEID, an extreme precision Doppler spectrograph for WIYN},'' in [{\em Ground-based and Airborne Instrumentation for Astronomy VI}{\nolinebreak\hspace{0.1em}]},  {Evans}, C.~J., {Simard}, L., and {Takami}, H., eds., {\em Society of Photo-Optical Instrumentation Engineers (SPIE) Conference Series} {\bf 9908},  99087H (Aug. 2016).

\bibitem{seifahrt2018}
{Seifahrt}, A., {St{\"u}rmer}, J., {Bean}, J.~L., and {Schwab}, C., ``{MAROON-X: a radial velocity spectrograph for the Gemini Observatory},'' in [{\em Ground-based and Airborne Instrumentation for Astronomy VII}{\nolinebreak\hspace{0.1em}]},  {Evans}, C.~J., {Simard}, L., and {Takami}, H., eds., {\em Society of Photo-Optical Instrumentation Engineers (SPIE) Conference Series} {\bf 10702},  107026D (July 2018).

\bibitem{pepe2021}
{Pepe}, F., {Cristiani}, S., {Rebolo}, R., {Santos}, N.~C., {Dekker}, H., {Cabral}, A., {Di Marcantonio}, P., {Figueira}, P., {Lo Curto}, G., {Lovis}, C., {Mayor}, M., {M{\'e}gevand}, D., {Molaro}, P., {Riva}, M., {Zapatero Osorio}, M.~R., {Amate}, M., {Manescau}, A., {Pasquini}, L., {Zerbi}, F.~M., {Adibekyan}, V., {Abreu}, M., {Affolter}, M., {Alibert}, Y., {Aliverti}, M., {Allart}, R., {Allende Prieto}, C., {{\'A}lvarez}, D., {Alves}, D., {Avila}, G., {Baldini}, V., {Bandy}, T., {Barros}, S.~C.~C., {Benz}, W., {Bianco}, A., {Borsa}, F., {Bourrier}, V., {Bouchy}, F., {Broeg}, C., {Calderone}, G., {Cirami}, R., {Coelho}, J., {Conconi}, P., {Coretti}, I., {Cumani}, C., {Cupani}, G., {D'Odorico}, V., {Damasso}, M., {Deiries}, S., {Delabre}, B., {Demangeon}, O.~D.~S., {Dumusque}, X., {Ehrenreich}, D., {Faria}, J.~P., {Fragoso}, A., {Genolet}, L., {Genoni}, M., {G{\'e}nova Santos}, R., {Gonz{\'a}lez Hern{\'a}ndez}, J.~I., {Hughes}, I., {Iwert}, O., {Kerber}, F., {Knudstrup}, J., {Landoni}, M., {Lavie}, B.,
  {Lillo-Box}, J., {Lizon}, J.~L., {Maire}, C., {Martins}, C.~J.~A.~P., {Mehner}, A., {Micela}, G., {Modigliani}, A., {Monteiro}, M.~A., {Monteiro}, M.~J.~P.~F.~G., {Moschetti}, M., {Murphy}, M.~T., {Nunes}, N., {Oggioni}, L., {Oliveira}, A., {Oshagh}, M., {Pall{\'e}}, E., {Pariani}, G., {Poretti}, E., {Rasilla}, J.~L., {Rebord{\~a}o}, J., {Redaelli}, E.~M., {Santana Tschudi}, S., {Santin}, P., {Santos}, P., {S{\'e}gransan}, D., {Schmidt}, T.~M., {Segovia}, A., {Sosnowska}, D., {Sozzetti}, A., {Sousa}, S.~G., {Span{\`o}}, P., {Su{\'a}rez Mascare{\~n}o}, A., {Tabernero}, H., {Tenegi}, F., {Udry}, S., and {Zanutta}, A., ``{ESPRESSO at VLT. On-sky performance and first results},'' {\em \aap}~{\bf 645},  A96 (Jan. 2021).

\bibitem{crass2021}
{Crass}, J., {Bechter}, A., {Sands}, B., {King}, D., {Ketterer}, R., {Engstrom}, M., {Hamper}, R., {Kopon}, D., {Smous}, J., {Crepp}, J.~R., {Montoya}, M., {Durney}, O., {Cavalieri}, D., {Reynolds}, R., {Vansickle}, M., {Onuma}, E., {Thomes}, J., {Mullin}, S., {Shelton}, C., {Wallace}, K., {Bechter}, E., {Vaz}, A., {Power}, J., {Rahmer}, G., and {Ertel}, S., ``{Final design and on-sky testing of the iLocater SX acquisition camera: broad-band single-mode fibre coupling},'' {\em \mnras}~{\bf 501},  2250--2267 (Feb. 2021).

\bibitem{bouchy2005}
{Bouchy}, F., {Bazot}, M., {Santos}, N.~C., {Vauclair}, S., and {Sosnowska}, D., ``{Asteroseismology of the planet-hosting star {\ensuremath{\mu}} Arae. I. The acoustic spectrum},'' {\em \aap}~{\bf 440},  609--614 (Sept. 2005).

\bibitem{kjeldsen2005}
{Kjeldsen}, H., {Bedding}, T.~R., {Butler}, R.~P., {Christensen-Dalsgaard}, J., {Kiss}, L.~L., {McCarthy}, C., {Marcy}, G.~W., {Tinney}, C.~G., and {Wright}, J.~T., ``{Solar-like Oscillations in {\ensuremath{\alpha}} Centauri B},'' {\em \apj}~{\bf 635},  1281--1290 (Dec. 2005).

\bibitem{arentoft2008}
{Arentoft}, T., {Kjeldsen}, H., {Bedding}, T.~R., {Bazot}, M., {Christensen-Dalsgaard}, J., {Dall}, T.~H., {Karoff}, C., {Carrier}, F., {Eggenberger}, P., {Sosnowska}, D., {Wittenmyer}, R.~A., {Endl}, M., {Metcalfe}, T.~S., {Hekker}, S., {Reffert}, S., {Butler}, R.~P., {Bruntt}, H., {Kiss}, L.~L., {O'Toole}, S.~J., {Kambe}, E., {Ando}, H., {Izumiura}, H., {Sato}, B., {Hartmann}, M., {Hatzes}, A., {Bouchy}, F., {Mosser}, B., {Appourchaux}, T., {Barban}, C., {Berthomieu}, G., {Garcia}, R.~A., {Michel}, E., {Provost}, J., {Turck-Chi{\`e}ze}, S., {Marti{\'c}}, M., {Lebrun}, J.-C., {Schmitt}, J., {Bertaux}, J.-L., {Bonanno}, A., {Benatti}, S., {Claudi}, R.~U., {Cosentino}, R., {Leccia}, S., {Frandsen}, S., {Brogaard}, K., {Glowienka}, L., {Grundahl}, F., and {Stempels}, E., ``{A Multisite Campaign to Measure Solar-like Oscillations in Procyon. I. Observations, Data Reduction, and Slow Variations},'' {\em \apj}~{\bf 687},  1180--1190 (Nov. 2008).

\bibitem{dumusque2011a}
{Dumusque}, X., {Udry}, S., {Lovis}, C., {Santos}, N.~C., and {Monteiro}, M.~J.~P.~F.~G., ``{Planetary detection limits taking into account stellar noise. I. Observational strategies to reduce stellar oscillation and granulation effects},'' {\em \aap}~{\bf 525},  A140 (Jan. 2011).

\bibitem{chaplin2019}
{Chaplin}, B., ``{Asteroseismology of planet hosts: summary and future challenges},'' in [{\em KITP Conference: Planet-Star Connections in the Era of TESS and Gaia}{\nolinebreak\hspace{0.1em}]},   9 (May 2019).

\bibitem{dravins1982}
{Dravins}, D., ``{Photospheric spectrum line asymmetries and wavelength shifts},'' {\em \araa}~{\bf 20},  61--89 (Jan. 1982).

\bibitem{kjeldsen1995}
{Kjeldsen}, H. and {Bedding}, T.~R., ``{Amplitudes of stellar oscillations: the implications for asteroseismology.},'' {\em \aap}~{\bf 293},  87--106 (Jan. 1995).

\bibitem{lindegreen2003}
{Lindegren}, L. and {Dravins}, D., ``{The fundamental definition of ``radial velocity''},'' {\em \aap}~{\bf 401},  1185--1201 (Apr. 2003).

\bibitem{meunier2015}
{Meunier}, N., {Lagrange}, A.~M., {Borgniet}, S., and {Rieutord}, M., ``{Using the Sun to estimate Earth-like planet detection capabilities. VI. Simulation of granulation and supergranulation radial velocity and photometric time series},'' {\em \aap}~{\bf 583},  A118 (Nov. 2015).

\bibitem{cegla2018}
{Cegla}, H.~M., {Watson}, C.~A., {Shelyag}, S., {Chaplin}, W.~J., {Davies}, G.~R., {Mathioudakis}, M., {Palumbo}, M.~L., I., {Saar}, S.~H., and {Haywood}, R.~D., ``{Stellar Surface Magneto-convection as a Source of Astrophysical Noise. II. Center-to-limb Parameterization of Absorption Line Profiles and Comparison to Observations},'' {\em \apj}~{\bf 866},  55 (Oct. 2018).

\bibitem{lanza2019}
{Lanza}, A.~F., {Gizon}, L., {Zaqarashvili}, T.~V., {Liang}, Z.~C., and {Rodenbeck}, K., ``{Sectoral r modes and periodic radial velocity variations of Sun-like stars},'' {\em \aap}~{\bf 623},  A50 (Mar. 2019).

\bibitem{rincon2018}
{Rincon}, F. and {Rieutord}, M., ``{The Sun's supergranulation},'' {\em Living Reviews in Solar Physics}~{\bf 15},  6 (Sept. 2018).

\bibitem{meunier2019}
{Meunier}, N. and {Lagrange}, A.~M., ``{Unexpectedly strong effect of supergranulation on the detectability of Earth twins orbiting Sun-like stars with radial velocities},'' {\em \aap}~{\bf 625},  L6 (May 2019).

\bibitem{saar1997}
{Saar}, S.~H. and {Donahue}, R.~A., ``{Activity-Related Radial Velocity Variation in Cool Stars},'' {\em \apj}~{\bf 485},  319--327 (Aug. 1997).

\bibitem{hatzes2002}
{Hatzes}, A.~P., ``{Starspots and exoplanets},'' {\em Astronomische Nachrichten}~{\bf 323},  392--394 (July 2002).

\bibitem{saar2003}
{Saar}, S.~H., ``{The Effects of Plage on Precision Radial Velocities},'' in [{\em Scientific Frontiers in Research on Extrasolar Planets}{\nolinebreak\hspace{0.1em}]},  {Deming}, D. and {Seager}, S., eds., {\em Astronomical Society of the Pacific Conference Series} {\bf 294},  65--70 (Jan. 2003).

\bibitem{desort2007}
{Desort}, M., {Lagrange}, A.~M., {Galland}, F., {Udry}, S., and {Mayor}, M., ``{Search for exoplanets with the radial-velocity technique: quantitative diagnostics of stellar activity},'' {\em \aap}~{\bf 473},  983--993 (Oct. 2007).

\bibitem{huelamo2008}
{Hu{\'e}lamo}, N., {Figueira}, P., {Bonfils}, X., {Santos}, N.~C., {Pepe}, F., {Gillon}, M., {Azevedo}, R., {Barman}, T., {Fern{\'a}ndez}, M., {di Folco}, E., {Guenther}, E.~W., {Lovis}, C., {Melo}, C.~H.~F., {Queloz}, D., and {Udry}, S., ``{TW Hydrae: evidence of stellar spots instead of a Hot Jupiter},'' {\em \aap}~{\bf 489},  L9--L13 (Oct. 2008).

\bibitem{dumusque2011b}
{Dumusque}, X., {Santos}, N.~C., {Udry}, S., {Lovis}, C., and {Bonfils}, X., ``{Planetary detection limits taking into account stellar noise. II. Effect of stellar spot groups on radial-velocities},'' {\em \aap}~{\bf 527},  A82 (Mar. 2011).

\bibitem{jeffers2013}
{Jeffers}, S.~V., {Barnes}, J.~R., {Jones}, H., and {Pinfield}, D., ``{Realistic limitations of detecting planets around young active stars},'' in [{\em European Physical Journal Web of Conferences}{\nolinebreak\hspace{0.1em}]},  {\em European Physical Journal Web of Conferences} {\bf 47},  09002 (Apr. 2013).

\bibitem{roettenbacher2022}
{Roettenbacher}, R.~M., {Cabot}, S. H.~C., {Fischer}, D.~A., {Monnier}, J.~D., {Henry}, G.~W., {Harmon}, R.~O., {Korhonen}, H., {Brewer}, J.~M., {Llama}, J., {Petersburg}, R.~R., {Zhao}, L.~L., {Kraus}, S., {Le Bouquin}, J.-B., {Anugu}, N., {Davies}, C.~L., {Gardner}, T., {Lanthermann}, C., {Schaefer}, G., {Setterholm}, B., {Clark}, C.~A., {Jorstad}, S.~G., {Kuehn}, K., and {Levine}, S., ``{EXPRES. III. Revealing the Stellar Activity Radial Velocity Signature of $\epsilon$ Eridani with Photometry and Interferometry},'' {\em \aj}~{\bf 163},  19 (Jan. 2022).

\bibitem{haywood2016}
{Haywood}, R.~D., {Collier Cameron}, A., {Unruh}, Y.~C., {Lovis}, C., {Lanza}, A.~F., {Llama}, J., {Deleuil}, M., {Fares}, R., {Gillon}, M., {Moutou}, C., {Pepe}, F., {Pollacco}, D., {Queloz}, D., and {S{\'e}gransan}, D., ``{The Sun as a planet-host star: proxies from SDO images for HARPS radial-velocity variations},'' {\em \mnras}~{\bf 457},  3637--3651 (Apr. 2016).

\bibitem{meunier2010}
{Meunier}, N., {Desort}, M., and {Lagrange}, A.~M., ``{Using the Sun to estimate Earth-like planets detection capabilities . II. Impact of plages},'' {\em \aap}~{\bf 512},  A39 (Mar. 2010).

\bibitem{egeland2017}
{Egeland}, R., {Soon}, W., {Baliunas}, S., {Hall}, J.~C., {Pevtsov}, A.~A., and {Bertello}, L., ``{The Mount Wilson Observatory S-index of the Sun},'' {\em \apj}~{\bf 835},  25 (Jan. 2017).

\bibitem{luhn2022}
{Luhn}, J.~K., {Wright}, J.~T., {Henry}, G.~W., {Saar}, S.~H., and {Baum}, A.~C., ``{HD 166620: Portrait of a Star Entering a Grand Magnetic Minimum},'' {\em \apjl}~{\bf 936},  L23 (Sept. 2022).

\bibitem{haywood2014}
{Haywood}, R.~D., {Collier Cameron}, A., {Queloz}, D., {Barros}, S.~C.~C., {Deleuil}, M., {Fares}, R., {Gillon}, M., {Lanza}, A.~F., {Lovis}, C., {Moutou}, C., {Pepe}, F., {Pollacco}, D., {Santerne}, A., {S{\'e}gransan}, D., and {Unruh}, Y.~C., ``{Planets and stellar activity: hide and seek in the CoRoT-7 system},'' {\em \mnras}~{\bf 443},  2517--2531 (Sept. 2014).

\bibitem{phillips2016}
{Phillips}, D.~F., {Glenday}, A.~G., {Dumusque}, X., {Buchschacher}, N., {Collier Cameron}, A., {Cecconi}, M., {Charbonneau}, D., {Cosentino}, R., {Ghedina}, A., {Haywood}, R., {Latham}, D.~W., {Li}, C.-H., {Lodi}, M., {Lovis}, C., {Molinari}, E., {Pepe}, F., {Sasselov}, D., {Szentgyorgyi}, A., {Udry}, S., and {Walsworth}, R.~L., ``{An astro-comb calibrated solar telescope to search for the radial velocity signature of Venus},'' in [{\em Advances in Optical and Mechanical Technologies for Telescopes and Instrumentation II}{\nolinebreak\hspace{0.1em}]},  {Navarro}, R. and {Burge}, J.~H., eds., {\em Society of Photo-Optical Instrumentation Engineers (SPIE) Conference Series} {\bf 9912},  99126Z (July 2016).

\bibitem{lin2022}
{Lin}, A. S.~J., {Monson}, A., {Mahadevan}, S., {Ninan}, J.~P., {Halverson}, S., {Nitroy}, C., {Bender}, C.~F., {Logsdon}, S.~E., {Kanodia}, S., {Terrien}, R.~C., {Roy}, A., {Luhn}, J.~K., {Gupta}, A.~F., {Ford}, E.~B., {Hearty}, F., {Laher}, R.~R., {Hunting}, E., {McBride}, W.~R., {Salazar Rivera}, N.~I., {Rajagopal}, J., {Wolf}, M.~J., {Robertson}, P., {Wright}, J.~T., {Blake}, C.~H., {Ca{\~n}as}, C.~I., {Lubar}, E., {McElwain}, M.~W., {Ramsey}, L.~W., {Schwab}, C., and {Stefansson}, G., ``{Observing the Sun as a Star: Design and Early Results from the NEID Solar Feed},'' {\em \aj}~{\bf 163},  184 (Apr. 2022).

\bibitem{claudi2018}
{Claudi}, R., {Ghedina}, A., {Pace}, E., {Gallorini}, L., {Di Giorgio}, A.~M., {Liu}, S.~J., {Tozzi}, A., {Lanza}, A.~F., {Micela}, G., {Molinari}, E., {Phillips}, D., and {Tripodo}, G., ``{LOCNES: low cost NIR extended solar telescope},'' in [{\em Ground-based and Airborne Telescopes VII}{\nolinebreak\hspace{0.1em}]},  {Marshall}, H.~K. and {Spyromilio}, J., eds., {\em Society of Photo-Optical Instrumentation Engineers (SPIE) Conference Series} {\bf 10700},  107004N (July 2018).

\bibitem{leite2022}
{Leite}, I., {Cabral}, A., {Abreu}, M., and {Santos}, N., ``{Imaging sensors for PoET: a spatially resolved solar spectroscopy instrument},'' in [{\em Journal of Physics Conference Series}{\nolinebreak\hspace{0.1em}]},  {\em Journal of Physics Conference Series} {\bf 2407},  012020, IOP (Dec. 2022).

\bibitem{rubenzahl2023}
{Rubenzahl}, R.~A., {Halverson}, S., {Walawender}, J., {Hill}, G.~M., {Howard}, A.~W., {Brown}, M., {Ida}, E., {Tehero}, J., {Fulton}, B.~J., {Gibson}, S.~R., {Kassis}, M., {Smith}, B., {Wold}, T., and {Payne}, J., ``{Staring at the Sun with the Keck Planet Finder: An Autonomous Solar Calibrator for High Signal-to-noise Sun-as-a-star Spectra},'' {\em \pasp}~{\bf 135},  125002 (Dec. 2023).

\bibitem{levine2012}
{Levine}, S.~E., {Bida}, T.~A., {Chylek}, T., {Collins}, P.~L., {DeGroff}, W.~T., {Dunham}, E.~W., {Lotz}, P.~J., {Venetiou}, A.~J., and {Zoonemat Kermani}, S., ``{Status and performance of the Discovery Channel Telescope during commissioning},'' in [{\em Ground-based and Airborne Telescopes IV}{\nolinebreak\hspace{0.1em}]},  {Stepp}, L.~M., {Gilmozzi}, R., and {Hall}, H.~J., eds., {\em Society of Photo-Optical Instrumentation Engineers (SPIE) Conference Series} {\bf 8444},  844419 (Sept. 2012).

\bibitem{brewer2020}
{Brewer}, J.~M., {Fischer}, D.~A., {Blackman}, R.~T., {Cabot}, S. H.~C., {Davis}, A.~B., {Laughlin}, G., {Leet}, C., {Ong}, J.~M.~J., {Petersburg}, R.~R., {Szymkowiak}, A.~E., {Zhao}, L.~L., {Henry}, G.~W., and {Llama}, J., ``{EXPRES. I. HD 3651 as an Ideal RV Benchmark},'' {\em \aj}~{\bf 160},  67 (Aug. 2020).

\bibitem{petersburg2020}
{Petersburg}, R.~R., {Ong}, J.~M.~J., {Zhao}, L.~L., {Blackman}, R.~T., {Brewer}, J.~M., {Buchhave}, L.~A., {Cabot}, S. H.~C., {Davis}, A.~B., {Jurgenson}, C.~A., {Leet}, C., {McCracken}, T.~M., {Sawyer}, D., {Sharov}, M., {Tronsgaard}, R., {Szymkowiak}, A.~E., and {Fischer}, D.~A., ``{An Extreme-precision Radial-velocity Pipeline: First Radial Velocities from EXPRES},'' {\em \aj}~{\bf 159},  187 (May 2020).

\bibitem{kanodia2018}
{Kanodia}, S. and {Wright}, J.~T., ``{Barycorrpy: Barycentric velocity calculation and leap second management}.'' Astrophysics Source Code Library, record ascl:1808.001 (Aug. 2018).

\bibitem{zhao2021}
{Zhao}, L.~L., {Hogg}, D.~W., {Bedell}, M., and {Fischer}, D.~A., ``{Excalibur: A Nonparametric, Hierarchical Wavelength Calibration Method for a Precision Spectrograph},'' {\em \aj}~{\bf 161},  80 (Feb. 2021).

\bibitem{leet2019}
{Leet}, C., {Fischer}, D.~A., and {Valenti}, J.~A., ``{Toward a Self-calibrating, Empirical, Light-weight Model for Tellurics in High-resolution Spectra},'' {\em \aj}~{\bf 157},  187 (May 2019).

\bibitem{colliercameron2019}
{Collier Cameron}, A., {Mortier}, A., {Phillips}, D., {Dumusque}, X., {Haywood}, R.~D., {Langellier}, N., {Watson}, C.~A., {Cegla}, H.~M., {Costes}, J., {Charbonneau}, D., {Coffinet}, A., {Latham}, D.~W., {Lopez-Morales}, M., {Malavolta}, L., {Maldonado}, J., {Micela}, G., {Milbourne}, T., {Molinari}, E., {Saar}, S.~H., {Thompson}, S., {Buchschacher}, N., {Cecconi}, M., {Cosentino}, R., {Ghedina}, A., {Glenday}, A., {Gonzalez}, M., {Li}, C.~H., {Lodi}, M., {Lovis}, C., {Pepe}, F., {Poretti}, E., {Rice}, K., {Sasselov}, D., {Sozzetti}, A., {Szentgyorgyi}, A., {Udry}, S., and {Walsworth}, R., ``{Three years of Sun-as-a-star radial-velocity observations on the approach to solar minimum},'' {\em \mnras}~{\bf 487},  1082--1100 (July 2019).

\bibitem{scherrer2012}
{Scherrer}, P.~H., {Schou}, J., {Bush}, R.~I., {Kosovichev}, A.~G., {Bogart}, R.~S., {Hoeksema}, J.~T., {Liu}, Y., {Duvall}, T.~L., {Zhao}, J., {Title}, A.~M., {Schrijver}, C.~J., {Tarbell}, T.~D., and {Tomczyk}, S., ``{The Helioseismic and Magnetic Imager (HMI) Investigation for the Solar Dynamics Observatory (SDO)},'' {\em \solphys}~{\bf 275},  207--227 (Jan. 2012).

\bibitem{pesnell2012}
{Pesnell}, W.~D., {Thompson}, B.~J., and {Chamberlin}, P.~C., ``{The Solar Dynamics Observatory (SDO)},'' {\em \solphys}~{\bf 275},  3--15 (Jan. 2012).

\bibitem{zhao2023}
{Zhao}, L.~L., {Dumusque}, X., {Ford}, E.~B., {Llama}, J., {Mortier}, A., {Bedell}, M., {Al Moulla}, K., {Bender}, C.~F., {Blake}, C.~H., {Brewer}, J.~M., {Collier Cameron}, A., {Cosentino}, R., {Figueira}, P., {Fischer}, D.~A., {Ghedina}, A., {Gonzalez}, M., {Halverson}, S., {Kanodia}, S., {Latham}, D.~W., {Lin}, A. S.~J., {Lo Curto}, G., {Lodi}, M., {Logsdon}, S.~E., {Lovis}, C., {Mahadevan}, S., {Monson}, A., {Ninan}, J.~P., {Pepe}, F., {Roettenbacher}, R.~M., {Roy}, A., {Santos}, N.~C., {Schwab}, C., {Stef{\'a}nsson}, G., {Szymkowiak}, A.~E., {Terrien}, R.~C., {Udry}, S., {Weiss}, S.~A., {Wildi}, F., {Wildi}, T., and {Wright}, J.~T., ``{The Extreme Stellar-signals Project. III. Combining Solar Data from HARPS, HARPS-N, EXPRES, and NEID},'' {\em \aj}~{\bf 166},  173 (Oct. 2023).

\end{thebibliography}
\bibliographystyle{spiebib} 

\end{document}